\AtBeginDvi{}
\documentclass[onecolumn, superscriptaddress]{jpsj3}
\usepackage[dvipdfmx]{color}
\usepackage{setspace}

\title{Importance of Fermi Surface Topology for In-Plane Resistivity Anisotropy in Hole- and Electron-Doped Ba(Fe$_{1-x}$TM$_{x}$)$_2$As$_2$ (TM=Cr, Mn and Co)}

\author{Tatsuya Kobayashi\thanks{kobayashi@tsurugi.phys.sci.osaka-u.ac.jp}, Kiyohisa Tanaka\thanks{Present address: UVSOR Facility, Institute for Molecular Science, Okazaki, Aichi 444-8585, Japan. School of Physical Sciences, The Graduate University for Advanced Studies (SOKENDAI), Okazaki, Aichi 444-8585, Japan}, Shigeki Miyasaka, Setsuko Tajima}
\inst{Department of Physics, Graduate School of Science, Osaka University, Toyonaka, Osaka 560-0043, Japan}

\abst{The in-plane anisotropy of resistivity has been investigated for Ba(Fe$_{1-x}$TM$_{x}$)$_2$As$_2$ (TM-Ba122, TM$=$Cr, Mn, and Co) where the substitution sites are the same but the doped carriers are different for different TM elements. 
The Hall coefficient measurements indicated that hole carriers are effectively doped by Cr substitution but not by Mn substitution.
It has been found that the resistivity difference $\Delta\rho=\rho_{\rm b}-\rho_{\rm a}$ in the antiferromagnetic-orthorhombic (AFO) phase of Cr-Ba122 is initially positive but it turns to negative with increasing Cr content, whereas the positive $\Delta\rho$ monotonically increases with Mn substitution in Mn-Ba122. 
In the paramagnetic-tetragonal phase, $\Delta\rho$ is always positive, but it decreases with substitution in Cr-Ba122, in contrast to the electron-doped case. 
    These results demonstrate that the resistivity anisotropy exhibits electron-hole asymmetry in both AFO and nematic phases and that it depends on the Fermi surface topology whether the carrier scattering results in a positive or negative $\Delta\rho$.}
 
\begin{document}
 
\maketitle
\section{Introduction}
      \par\par Since the discovery of superconductivity in iron pnictides, the superconducting and normal state properties of this system have been extensively investigated \cite{Hosono, Stewart}. 
      In almost all iron pnictides, the terminal compositions, BaFe$_2$As$_2$ and LaFeAsO, undergo a structural and magnetic phase transition from a higher-temperature paramagnetic-tetragonal (PT) state to a lower-temperature antiferromagnetic-orthorhombic (AFO) one with decreasing temperature. By substituting various elements or applying pressure, the AFO state is suppressed and superconductivity emerges. 
      One of the puzzles in the normal-state properties is the anomalous electronic anisotropy in the AFO and paramagnetic-orthorhombic phases \cite{Fisher, Fernandes}. 
      Although the anisotropy in the paramagnetic-orthorhombic phase, which is the so-called nematic phase \cite{Chu, Yi, Blomberg2, Yi2}, has attracted much interest in terms of the nematic nature of the electronic state, the anisotropy in the AFO phase is also unusual. 
      Particularly, the in-plane resistivity shows significant anisotropy in the AFO phase at low temperatures \cite{Chu}. 
      The resistivity along the longer $a$-axis with an antiferromagnetic spin arrangement ($\rho_{\rm a}$) is smaller than that along the shorter $b$-axis with a ferromagnetic spin arrangement ($\rho_{\rm b}$). 
      This observed anisotropy ($\rho_{\rm b}>\rho_{\rm a}$) is counterintuitive because both of the larger orbitals overlap due to a smaller lattice constant and smaller spin-fluctuation scattering due to a ferromagnetic spin arrangement, which should give a smaller resistivity in the $b$-direction ($\rho_{\rm b}<\rho_{\rm a}$). 
      Several theories based on magnetic order, orbital order, and lattice distortion have been proposed to explain this anomalous anisotropic resistivity \cite{Liang, Chen, Zhang, Lv, Laad, Liang2}. 
   However, all of them fail to explain the fact that transition metal (TM) substitution for the Fe sites enhances the resistivity anisotropy despite the suppression of the magnetic order, orbital polarization, and lattice orthorhombicity \cite{Kuo, Yi}. 
\begin{figure}
\begin{center}
\includegraphics[width=0.40\textwidth]{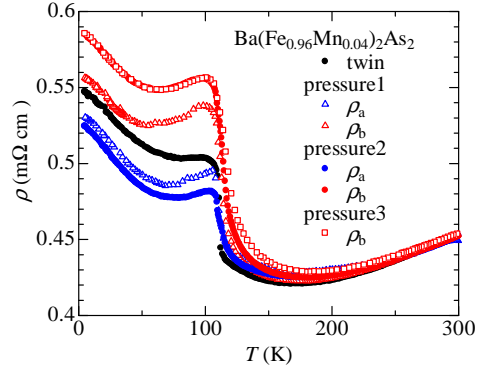}
\end{center}
\caption{(Color online) Temperature ($T$) dependence of the in-plane resistivity along $a$-axis ($\rho_{\rm a}$) (blue) and $b$-axis ($\rho_{\rm b}$) (red) of detwinned Ba(Fe$_{0.96}$Mn$_{0.04}$)$_2$As$_2$ under different pressures. Triangles indicate the result under the first pressure (Pressure1) and circles and squares under larger pressures (Pressure2 and Pressure3), respectively. 
The largest pressure (Pressure3) data almost completely overlap the second largest pressure (Pressure2) data, suggesting saturation of the effect due to complete detwinning.}
\label{Fig5}
\end{figure}

\begin{figure*}
\begin{center}
\includegraphics[width=0.9\textwidth]{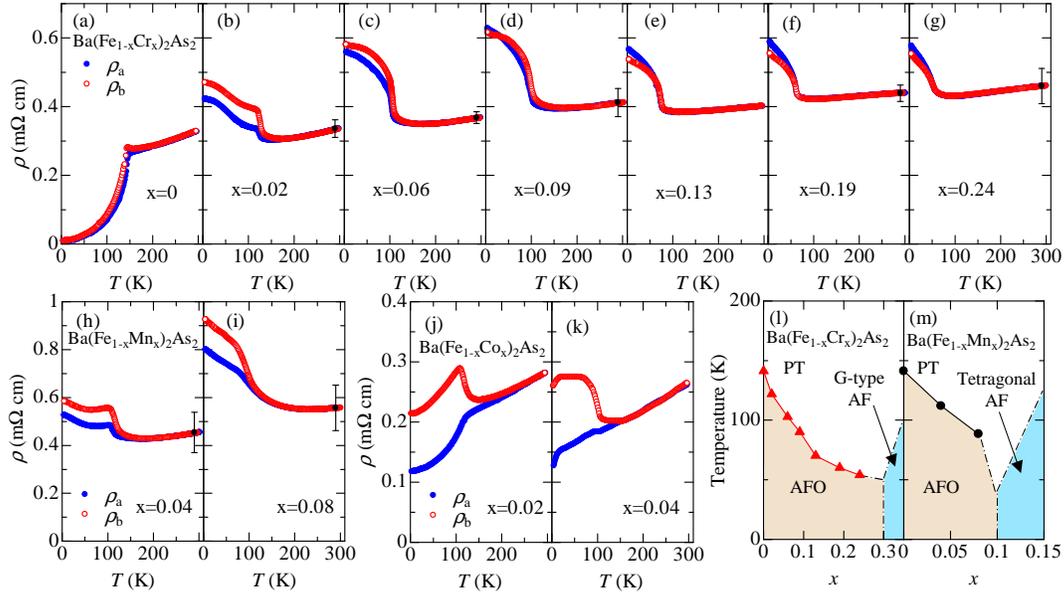}\\
\end{center}
\caption{(Color online) Temperature ($T$) dependence of the in-plane resistivity along the $a$-axis ($\rho_{\rm a}$) (blue, closed circles) and $b$-axis ($\rho_{\rm b}$) (red, open circles) of detwinned Ba(Fe$_{1-x}$TM$_{x}$)$_2$As$_2$ {TM=Cr [(a)-(g)], Mn [(h), (i)] and Co [(j), (k)]}.
Statistical error bars of quantitative resistivity values are shown for Ba(Fe$_{1-x}$TM$_{x}$)$_2$As$_2$ (TM=Cr and Mn). 
(l) and (m) show the $T-x$ phase diagram of Ba(Fe$_{1-x}$Cr$_{x}$)$_2$As$_2$ and Ba(Fe$_{1-x}$Mn$_{x}$)$_2$As$_2$, respectively. 
The magnetostructural transition temperatures, $T_\mathrm{AFO}$, determined from $d \rho (T)/dT$ with twinned crystals, are indicated by triangles in (l) and circles in (m). The Cr-doped system with $x\geq0.30$ and the Mn-doped one with $x\geq0.10$ show the magnetic transition to a G-type antiferromagnetic (AF) state and an AF-tetragonal structural state, respectively, as reported in previous studies \cite{Marty, Kim}.}
\label{Fig1}
\end{figure*}
      To understand these counterintuitive behaviors, two different mechanisms have been proposed. 
   One is based on the anisotropy of impurity scattering and the other on the effect of anisotropic Fermi surface topology.
  The key experimental fact for the first scenario is that the anisotropy of in-plane resistivity in BaFe$_2$As$_2$ (Ba122) almost disappears after the postannealing treatment, which removes defects in the crystals \cite{Ishida, Ishida2}. 
  It was also pointed out that the anisotropy decreases with increasing distance of the substitution site from the Fe sites \cite{Ishida3}. 
  Based on these results, Ishida and co-workers proposed that the anisotropy of impurity scattering by substituted atoms is the origin of the anisotropy of in-plane resistivity, in other words, that the observed anisotropy is an extrinsic property. 
  This scenario is supported by optical \cite{Nakajima, Nakajima2} and scanning tunnel spectroscopy studies \cite{Chuang, Allan, Rosenthal}, as well as by theoretical investigations \cite{Gastiasoro, Kang, Inoue}. 
    
      Another important experimental result for the second scenario is the comparative study of electron-doped Ba(Fe$_{1-x}$Co$_{x}$)$_2$As$_2$ (Co-Ba122) and hole-doped (Ba$_{1-x}$K$_x$)Fe$_2$As$_2$ (K-Ba122). 
      The difference of resistivity ($\Delta\rho=\rho_{\rm b}-\rho_{\rm a}$) is smaller in K-Ba122 than in Co-Ba122, and more surprisingly, it becomes negative ($\rho_{\rm b}<\rho_{\rm a}$) with further K substitution \cite{Ying, Blomberg}. 
  Theoretically, the reversal of anisotropy has been predicted by considering the change of the Fermi surface topology and Drude weight in the AFO phase \cite{Blomberg, Valenzuela}. 
  However, they do not explain the postannealing effect, and recently it has been pointed out that the calculated anisotropy of the Drude weight is inconsistent with the experimental results \cite{Sugimoto}. 
  The mechanism based on spin-fluctuation scattering also takes into account the topology of the Fermi surface \cite{Fernandes2, Breitkreiz}, but it is applicable only in the nematic phase, not in the AFO phase. 
  In addition, some studies have regarded the observed inverse anisotropy as negligibly small, which is due to weak carrier scattering because of the off-site substitution \cite{Ishida3, Gastiasoro, Inoue}. 
  Thus, the previous studies could not distinguish the two factors of the in-plane resistivity anisotropy because they compared different site-substituted systems, and it turns out that the origin of the in-plane resistivity anisotropy is still controversial. 
  
    To distinguish the roles of the impurity scattering and the band structure, a comparative study using materials in which holes and electrons are doped by chemical substitution at the same atomic sites should be performed.  
    In the present study, we chose Ba(Fe$_{1-x}$TM$_{x}$)$_2$As$_2$ (TM-Ba122, TM=Cr, Mn) as the hole-doped systems. 
  Even when holes are doped, they do not show superconductivity and have the same AFO phase in the underdoped region ($x\leqq0.30$ for Cr and $x\leqq0.10$ for Mn) as in doped Ba122 \cite{Marty, Kim}. 
  (In the higher-doped region, Cr- and Mn-Ba122 systems show different types of antiferromagnetic order.)
  Since Cr, Mn, and Co are substituted for Fe, hole-doped TM-Ba122 (TM=Cr and Mn) is a suitable counterpart of the electron-doped Co-Ba122 to discuss the role of impurity scattering and band structure. 
  The purpose of the present work is to clarify the role of the impurity scattering and the topology of the Fermi surface in the in-plane resistivity anisotropy in the AFO and nematic phases of iron pnictide superconductors.


\begin{figure}
\begin{center}
\includegraphics[width=0.40\textwidth]{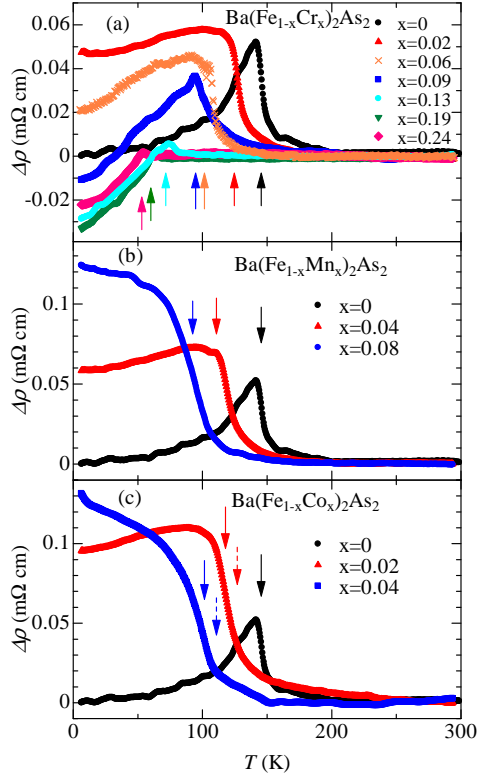}
\end{center}
\caption{(Color online) Temperature dependence of resistivity difference $\Delta\rho=\rho_{\rm b}-\rho_{\rm a}$ for (a) Ba(Fe$_{1-x}$Cr$_{x}$)$_2$As$_2$ with $x=0-0.24$, (b) Ba(Fe$_{1-x}$Mn$_{x}$)$_2$As$_2$ with $x=0-0.08$, and (c) Ba(Fe$_{1-x}$Co$_{x}$)$_2$As$_2$ with $x=0-0.04$. The arrows in (a) and (b) represent the magnetostructural transition temperature, $T_\mathrm{AFO}$. The solid and dotted arrows in (c) represent the magnetic and structural transition temperatures, respectively. The magnetic transition coincides with the structural transition in Cr- and Mn-Ba122, whereas the two transitions separate with substitution in Co-Ba122 \cite{Marty, Kim}.}
\label{Fig2}
\end{figure}
\section{Experimental Methods and Sample Characterization}
 Single crystals of Ba(Fe$_{1-x}$TM$_{x}$)$_2$As$_2$ (TM=Cr, Mn, and Co) were grown by a self-fluxing method \cite{Sefat, Thaler, Ishida2}. 
 Ba, FeAs, and TMAs were mixed in the atomic ratio 1:4($1-x$):$4x$, placed in an alumina crucible, and sealed in a quartz tube. 
 The tube was heated to 1200\,$^\circ$C, kept at that temperature for 10\,h, and cooled to 1000\,$^\circ$C at a rate of 2\,$^\circ$C/h. 
The compositions of the grown crystals were determined by scanning electron microscopy-eneregy dispersion X-ray (SEM-EDX) analysis. 
The crystals with $x>0.24$ for Cr and $x>0.08$ for Mn were not measured because they showed different magnetic and crystal structures from Co-Ba122 \cite{Marty, Kim}.
All the crystals were sealed in an evacuated silica tube and postannealed for several days at 800\,$^\circ$C. 
The crystals were detwinned by applying mechanical uniaxial pressure with a device similar to the previously reported one \cite{Chu}. 
In the AFO phase, the shorter $b$-axis is naturally aligned in the direction of the applied uniaxial pressure.
The measurements of the in-plane resistivity along both $a$- and $b$-axes of the orthorhombic lattice were performed on detwinned samples by a standard four-probe method. 
We repeated the resistivity measurements with increasing pressure and confirmed the saturation of the anisotropy, as shown in Fig. 1.
The magnetic field dependence of the Hall resistivity, $\rho_\mathrm{xy}$, was measured in magnetic fields up to 7\,T at various temperatures using twinned crystals. 
The Hall coefficient, $R_\mathrm{H}$, was determined by the polynomial fitting of the Hall resistivity with $\rho_\mathrm{xy}=R_\mathrm{H}H+aH^3$, where $H$ is the magnetic field.
\section{Results and Discussion}
       Figures 2(a)-(k) show the temperature dependence of the in-plane resistivity for detwinned TM-Ba122 (TM=Cr, Mn, and Co). 
All the crystals of Cr- and Mn-Ba122 measured show features consistent with the AFO transition, as in Co-Ba122 \cite{Marty, Kim}. 
Further increases of Cr and Mn content led to other ordered phases, as shown in Figs. 2(l) and 2(m). In this study, we investigated the resistivity anisotropy in the composition region of $x\leqq0.24$ for Cr and $x\leqq0.08$ for Mn.

	At room temperature, the resistivity monotonically increases with increasing Cr and Mn content beyond the statistical error, in contrast to the case of Co- and K-Ba122 \cite{Ishida3}. 
This suggests that Cr and Mn atoms in the FeAs plane are stronger scatterers than Co atoms. 
The resistivity decreases with decreasing temperature down to the magnetostructural transition temperature, $T_\mathrm{AFO}$. 
Below $T_\mathrm{AFO}$, the resistivity increases with decreasing temperature in both Cr- and Mn-Ba122. 
For $x=0$ [Fig. 2(a)], resistivity anisotropy between $\rho_{\rm a}$ and $\rho_{\rm b}$ is observed only around $T_\mathrm{AFO}$, as previously reported \cite{Ishida}. 
For $x=0.02$ of Cr-Ba122 [Fig. 2(b)], $\rho_{\rm b}$ is larger than $\rho_{\rm a}$ below $T_\mathrm{AFO}$, which is similar to the electron-doped Co-Ba122 shown in Figs. 2(j) and 2(k). 
Remarkably, the anisotropy decreases with further Cr substitution, and eventually one can see the crossover from $\rho_{\rm a} < \rho_{\rm b}$ to $\rho_{\rm a} > \rho_{\rm b}$ at low temperatures above $x=0.09$ [Figs. 2(d)-(g)]. 
On the other hand, in the case of Mn substitution, a clear anisotropic feature of $\rho_{\rm a} < \rho_{\rm b}$ continues up to $x=0.08$, as shown in Figs. 2(h) and 2(i). 


      We summarize the temperature dependence of the in-plane resistivity anisotropy $\Delta\rho(T)=\rho_{\rm b}-\rho_{\rm a}$ for TM-Ba122 (TM=Cr, Mn, and Co) in Fig. 3. 
When the temperature decreases, $\Delta\rho(T)$ starts to increase above $T_\mathrm{AFO}$ in all the samples, suggesting that the nematic phase is induced by applying pressure in this temperature range \cite{Blomberg2, Yi2}. 
For Cr-Ba122, $\Delta\rho(T)$ at $x=0.02$ is larger than that at $x=0$, as shown in Fig. 3(a). 
With further Cr substitution, a maximum $\Delta\rho(T)$ around $T_\mathrm{AFO}$ decreases but remains positive, whereas $\Delta\rho(T)$ at the lowest temperature has a negative value above $x=0.09$. 
As a result, an intersection of $\rho_{\rm a}$ and $\rho_{\rm b}$, namely, $\Delta\rho = 0$, is observed between $T_\mathrm{AFO}$ and the lowest temperature. 
This is significantly different behavior from that of Mn- and Co-Ba122, as shown in Figs. 3(b) and 3(c). 
In both systems, $T_\mathrm{AFO}$ is suppressed by the Mn and Co substitution, whereas the anisotropy of the in-plane resistivity, $\Delta\rho$, at the lowest temperature monotonically increases with $x$ below $x=0.08$ in Mn-Ba122 and below $x=0.04$ in Co-Ba122. In Mn- and Co-Ba122, $\Delta\rho(T)$ is always positive in all the temperature and composition ranges.
 
      To clarify the origin of the difference between Cr-Ba122 and Mn-Ba122, we performed Hall resistivity measurements. 
Figures 4(a) and 4(b) show the temperature dependence of the Hall coefficient, $R_\mathrm{H}(T)$, of Cr- and Mn-Ba122, respectively. 
Above $T_\mathrm{AFO}$, the sign of $R_\mathrm{H}(T)$ changes from negative to positive with Cr substitution, whereas $R_\mathrm{H}(T)$ slightly changes but remains negative with Mn substitution. 
This suggests that holes are doped by Cr substitution but not by Mn substitution, which will be more clearly discussed later. 
Around $T_\mathrm{AFO}$, $|R_\mathrm{H}(T)|$  abruptly increases due to the reconstruction of the Fermi surface. 
It is noted that $R_\mathrm{H}(T)$ below $T_\mathrm{AFO}$ shows a local minimum or maximum in Cr- and Mn-Ba122, which is different from K-Ba122 \cite{Shen, Ohgushi}. The origin of these temperature dependences may be related to the multi-band effect \cite{Albenque, Breitkreiz2}.


       Figure 5 shows the doping dependence of $R_\mathrm{H}$ at $T_\mathrm{AFO}$ and $\Delta\rho=\rho_{\rm b}-\rho_{\rm a}$ at 5\,K. 
At $T_\mathrm{AFO}$, $R_\mathrm{H}$ of Cr-Ba122 increases with $x$, resulting in a sign change around $x=0.09$, and almost saturates above $x=0.13$, while $R_\mathrm{H}$ of Mn-Ba122 decreases with doping and remains negative. 
A similar peak structure of $R_\mathrm{H}$ around $T_\mathrm{AFO}$ with doping is observed in the hole-doped K-Ba122 \cite{Shen, Ohgushi}. 
This suggests that the holes are effectively doped into Cr-Ba122 but not into Mn-Ba122. 
The almost absence of carrier doping in Mn-Ba122 is also suggested by the nuclear magnetic resonance and photoemission measurements \cite{Texier, Suzuki}. 

As shown in Fig. 5(b), $\Delta\rho$(5\,K) for Cr-Ba122 increases with increasing $x$ up to $x=0.02$ and then decreases above this composition. 
As a result, $\Delta\rho$(5\,K) of Cr-Ba122 shows a sign change at approximately $x=0.09$. 
Above $x=0.13$, it becomes nearly doping independent, corresponding to the saturation of $R_\mathrm{H}$($T_\mathrm{AFO}$). 
In contrast to Cr-Ba122, $\Delta\rho$(5\,K) of Mn-Ba122 monotonically increases with increasing doping. 

There are several possible origins of the difference in $\Delta\rho(T)$ in the AFO phase among Cr-Ba122, Mn-Ba122, and Co-Ba122. 
One is the crystallographic effect, namely, the change of lattice constants with doping affects $\Delta\rho(T)$. 
This is unlikely, however, because of the following reason. 
According to previous studies, the lattice constants of $a$- and $c$-axes increase with increasing Cr and/or Mn substitution in Ba122 \cite{Sefat, Thaler}, whereas they decrease with increasing Co substitution \cite{Ni}. 
On the other hand, $\Delta\rho(T)$ is similar in Mn-Ba122 and Co-Ba122 but different between Mn-Ba122 and Cr-Ba122. 
Therefore, there is no correlation between crystallographic change and $\Delta\rho(T)$ in these three systems. 
       
The second possible origin is the difference in carrier scattering. 
Above $T_\mathrm{AFO}$, $\rho(T)$ of Cr-, Mn-, Co-, and K-Ba122 \cite{Ishida3} is more or less similar. 
Below $T_\mathrm{AFO}$, however, $\rho(T)$ of Cr-, Mn-, and Co-Ba122 increases upon cooling , whereas $\rho(T)$ of K-Ba122 decreases.
This difference originates from the difference in the impurity scattering strength, which depends on the substituted element and site. 
Nevertheless, the difference in carrier scattering cannot explain the difference of $\Delta\rho(T)$. 
One example is that $\Delta\rho(T)$ of Mn-Ba122 is similar to that of Co-Ba122, despite the different $\rho(T)$, namely, different impurity scattering levels. 
A similar relation is observed between Cr-Ba122 and K-Ba122. These two compounds show different $\rho(T)$ but similar $\Delta\rho(T)$.
Therefore, the impurity scattering strength alone does not determine $\Delta\rho(T)$ in iron pnictides.  

The third possibility is the different carrier doping level. 
The present results can be summarized as follows. 
The resistivity anisotropy, $\Delta\rho$, in the AFO phase is relatively small and shows a sign change whenever hole carriers are doped into the system, irrespective of the chemical substitution site. 
This indicates that the size and shape of Fermi surfaces strongly affect the in-plane resistivity anisotropy. 
Thus, the theories which attribute the small $\Delta\rho$ in K-Ba122 to the absence of strong disorder \cite{Gastiasoro, Kang, Inoue} are inadequate as a general explanation of the in-plane resistivity anisotropy in iron pnictides.

Of course, we cannot ignore the impurity scattering effect, considering the annealing effect \cite{Ishida, Ishida2}. 
Actually, it has been pointed out that the theoretical calculation of conductivity based on the Fermi surface topology alone predicts anisotropy opposite to the observed one \cite{Sugimoto}. 
Therefore, it is likely that the anisotropy of resistivity of Ba122 systems in the AFO phase is induced by the anisotropic impurity scattering that reflects the anisotropic electronic state (Fermi surface), as recently proposed \cite{Sugimoto}. 
This scenario can explain not only the large anisotropy in the electron-doped Co-Ba122, where both electron doping and impurity scattering result in a positive $\Delta\rho$, but also that of Mn-Ba122, where the disorder determines $\Delta\rho$ because the Fermi surface does not change very much due to the absence of carrier doping. 
In the hole-doped systems, the resistivity anisotropy is intrinsically small and shows a negative $\Delta\rho$, regardless of the impurity scattering strength.


\begin{figure}
\begin{center}
\includegraphics[width=0.40\textwidth]{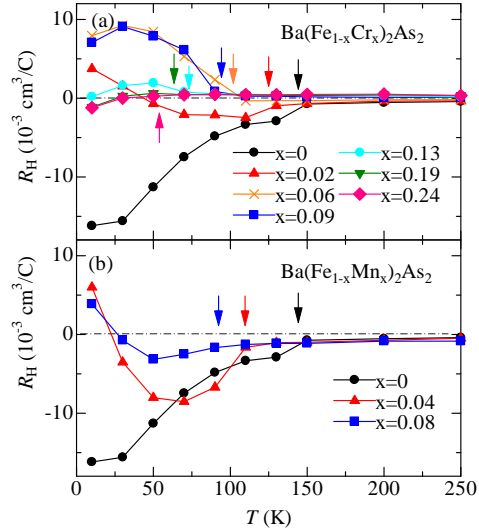}
\end{center}
\caption{(Color online) Temperature dependence of Hall coefficient $R_\mathrm{H}(T)$ for (a) Ba(Fe$_{1-x}$Cr$_{x}$)$_2$As$_2$ and (b) Ba(Fe$_{1-x}$Mn$_{x}$)$_2$As$_2$. The arrows indicate the position of $T_\mathrm{AFO}$.}
\label{Fig3}
\end{figure}

      Finally, we discuss the anisotropy at $T>T_\mathrm{AFO}$. 
      We can consider the different origin of the resistivity anisotropy below and above $T_\mathrm{AFO}$ because the AFO and nematic phases have different electronic structures \cite{Yi, Yi2}. 
       A small positive $\Delta\rho$($T_\mathrm{AFO}$) decreases with increasing Cr content in Cr-Ba122, as shown in Fig. 5(b). 
       This is similar to the results for K-Ba122 and Na-substituted CaFe$_2$As$_2$, where a very small $\Delta\rho$ is observed above $T_\mathrm{AFO}$ \cite{Ying, Blomberg, Ma}, but different from the results for the electron-doped Co-Ba122, where a positive $\Delta\rho$ is enhanced with doping \cite{Chu, Blomberg}. 
       Here again, the behaviors of $\Delta\rho$ are similar among the hole-doped systems but different between the hole- and electron-doped systems, irrespective of the substitution site. 
Therefore, the impurity scattering caused by the substituted atoms does not play a major role, instead, the change of the Fermi surface by carrier doping is crucial in the resistivity anisotropy above $T_\mathrm{AFO}$. 

Several theories have been proposed to explain the anisotropy in the nematic phase, such as mechanisms based on impurity scattering with an orbital order \cite{Inoue, Sugimoto} or emergent defect states \cite{Gastiasoro2}. 
We cannot support these scenarios based on impurity scattering, however, because it does not explain the observed electoron-hole asymmetry in $\Delta\rho$.
On the other hand, the spin-fluctuation scattering mechanism \cite{Fernandes2, Breitkreiz} predicts the electron-hole asymmetry of the anisotropy depending on the topology of the Fermi surface in the nematic phase. 
Because the present results can be explained by this mechanism, the spin-fluctuation scattering mechanism would be a strong candidate for the theory in the nematic phase. 
Recently, Kuo $et\ al.$ \cite{Kuo2} reported that the anisotropy in the nematic phase of the electron-doped Ba122 is independent of disorder, which is consistent with our conclusion.
\begin{figure}
 \begin{center}
\includegraphics[width=0.40\textwidth]{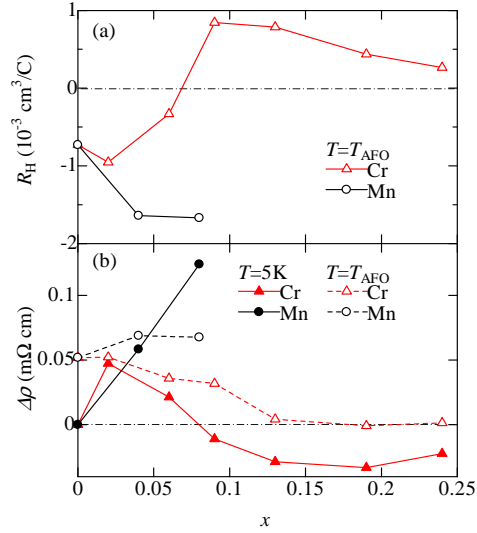}
\end{center}
\caption{(Color online) The doping ($x$) dependence of (a) Hall coefficient $R_\mathrm{H}$ at $T_\mathrm{AFO}$ (open symbols) and (b) $\Delta\rho=\rho_{\rm b}-\rho_{\rm a}$ at 5\,K (closed ones) and $T_\mathrm{AFO}$ (open ones) for Ba(Fe$_{1-x}$TM$_{x}$)$_2$As$_2$ [TM=Cr (red triangles) and Mn (black circles)].}
\label{Fig4}
\end{figure}

\section{Conclusion}
       We found that the in-plane resistivity anisotropy, $\Delta\rho$, in the AFO phase is small, but it clearly shows a sign change with increasing substituent and decreasing temperature in hole-doped Cr-Ba122. This is similar to the case of hole-doped K-Ba122 but different from the cases of the electron-doped Co-Ba122 and Mn-Ba122 where holes are almost undoped. 
       Moreover, the anisotropy above $T>T_\mathrm{AFO}$ shows electron-hole asymmetry in Co- and Cr-Ba122.
       These results demonstrate that the doping-dependent anisotropy of the Fermi surface indeed plays a dominant role in the resistivity anisotropy in both the AFO and the nematic phases, irrespective of the chemical substitution sites. 
Our results suggest that all the proposed mechanisms that attribute the electron-hole asymmetry of $\Delta\rho$ to the difference in the impurity scattering strength alone are inadequate. 
Rather, it depends on the electron-hole asymmetry of Fermi surface topology whether the impurity scattering results in a positive or negative $\Delta\rho$.

\begin{acknowledgments}
The authors appreciate T. Tohyama and M. Nakajima for helpful discussions. They also thank M. Ashida and M. Ichimiya for the technical support of SEM-EDX measurements. T. K. acknowledges the Grant-in-Aid for JSPS Fellows. The present work was supported by IRON-SEA, JST.
\end{acknowledgments}



\end{document}